\documentstyle[aas2pp4]{article}

\newcommand{\beq}{\begin{equation}}
\newcommand{\eeq}{\end{equation}}

\def\beqa{\begin{eqnarray}}
\def\eeqa{\end{eqnarray}}
\newcommand{\lsim}{\lesssim}
\newcommand{\gsim}{\gtrsim}
\def\p{\partial}
\def\lb{\langle}
\def\rb{\rangle}

\begin{document}
\twocolumn[
\title{Fluid Dark Matter}

\author{P. J. E. Peebles}
\affil{Joseph Henry Laboratories, Princeton University,
Princeton, NJ 08544}
 
\begin{abstract}
Dark matter modeled as a classical scalar field that interacts
only with gravity and with itself by a potential that is close to
quartic at large field values and approaches a quadratic form
when the field is small would be gravitationally produced by
inflation and at the present epoch could act like an ideal fluid
with pressure that is a function only of the mass density. 
This could have observationally interesting effects on the core
radii and solid body rotation of dark matter halos and on the
low mass end of the primeval mass fluctuation power spectrum. 
\end{abstract}
\vspace{1 mm}

\vspace{3 mm}\
]

\section{Introduction}
This paper continues the discussion by Peebles \&\
Vilenkin~(1999, hereafter PV) of a model for dark matter as a
gravitationally produced self-interacting scalar field. 
It is shown that this dark matter acts like an ideal fluid with
pressure that is a function of density alone, with possibly 
interesting consequences for the structure of dark halos of
galaxies at low redshift and the evolution of the
dark matter distribution at redshift $z\sim 10^4$ out of which
the galaxies are thought to have
formed. These effects are discussed in \S 2, and the
fluid model derived in \S 3. The distinction between the
behavior of this fluid and a gas of interacting dark 
matter particles (Spergel \&\ Steinhardt 1999) is discussed 
in \S 4, along with the possible significance of the fluid
picture.  

The model is motivated by the following
considerations. Fields that interact only with gravity would be
gravitationally produced by inflation (Ford 1987), and a
gravitationally produced scalar field would be left in a squeezed
state (Grishchuk \&\ Sidorov 1990) that acts like a classical
field. One could imagine that during inflation the 
potential of a scalar field that will end up as dark 
matter is close to $V=\lambda y^4/4$, and that inflation is
driven by a field $\phi$ with another quartic potential, 
$U=\lambda_\phi \phi ^4/4$. If the dimensionless
coefficients satisfy (PV eqs.~[85] and~[86]; for discussions of
other aspects of this constraint see Kofman \& Linde 1987; 
Felder, Kofman \&\ Linde 1999; and Starobinsky \&\ Yokoyama 1994)
\beq
10^{-6}\gsim\lambda > \lambda _\phi\lsim 10^{-14}, 
\label{eq:lambda}
\eeq
the mass density $\rho _y$ in dark matter at the end of
inflation may be an interesting fraction of the mass density
that becomes ordinary matter and radiation, and the space
distribution of $\rho _y$ is smooth enough that it does not
violate the isotropy of the thermal background radiation.
When $V(y)$ is close to quartic the dark mass 
density after inflation varies as $\rho _y\propto a(t)^{-4}$,
where $a(t)$ is the expansion parameter, meaning $\rho _y$ is
a fixed fraction of the mass density in radiation. Since the
field $y$ is supposed to have low pressure now the
curvature of the potential has to become close to quadratic at
small $|y|$. I adopt the form 
\beq
V=m^2y^2/2 + K|y|^q/q ,\qquad q>2,
\label{eq:V(y)}
\eeq
for typical field values at redshifts $z\lsim 10^5$.  
PV take $q=4$, making $K$ dimensionless. 
The extra parameter $q$ broadens the possibilities 
for fitting to the astronomy.

The analysis in \S 3 shows that 
when the anharmonic term in equation~(\ref{eq:V(y)}) is
subdominant the $y$-field mass behaves like an ideal
nonrelativistic fluid with
density $\rho _y=m^2\lb y^2\rb$ and pressure   
$p_y = F\rho _y{}^{q/2}$, where 
\beq
F = {q-2\over q}{2^{q/2-1}K\over\pi ^{1/2}m^q}
{\Gamma ((q+1)/2)\over\Gamma ((q+2)/2)}.
\label{eq:p_y}
\eeq
The velocity of sound, $c_s=(dp_y/d\rho _y)^{1/2}$, multiplied by 
the gravitational time $(G\rho _y)^{-1/2}$, is a Jeans
length $\propto\rho _y^{q/4-1}$.

\section{The Jeans Length}

In a pressure-supported dark matter halo the mass density as a
function of radius $r$ near the center is 
\beq
\rho _y(r) = \rho _c(1 - r^2/r_c^2),\quad
r_c=\left( 3c_s^2\over 2\pi G\rho _c\right) ^{1/2}.
\label{eq:rhoc}
\eeq
The rotation curve at
$r\lsim r_c$ is the solid body form 
\beq
v(r) = v_cr/r_c,\qquad v_c=2^{1/2}c_s .
\label{eq:sb}
\eeq
Since the halo can be supported by
Reynolds stress as well as pressure the condition on the size
of the region of solid body rotation in a halo of fluid dark matter
with central density $\rho _c$ is
\beq
r_c\geq R\rho _c^{q/4 - 1}.
\label{eq:r_c}
\eeq
for some constant, $R$.

A second constraint follows from the evolution 
of the fluctuation spectrum of the $y$-mass distribution through
the end of radiation-dominated expansion at redshift 
$z_{\rm eq}=2.4\times 10^4\Omega _yh^2$ and world time 
$t_{\rm eq}\sim 3\times 10^{10}\Omega _y^{-2}h^{-4}$~s,
for density parameter $\Omega _y$ in dark matter and  
Hubble parameter $h$ in units of
100~km~s$^{-1}$~Mpc$^{-1}$. The comoving wavenumber $k_c$ (scaled
to physical length at the present epoch) at the first zero of the
transfer function produced by the $y$-field pressure 
satisfies, in order of magnitude (Peebles 1980, eq.~[92.27]),
$k_c c_st_{\rm eq}z_{\rm eq}\sim\pi$.
At $k\gsim k_c$ the transfer function is 
suppressed relative to pressureless dark matter. A 
lower bound on $k_c$ for an acceptable model of structure
formation yields an upper bound on $c_s$ at
$z\sim z_{\rm eq}$.  

At the bounds on $q$ from $k_c$ and equation~(\ref{eq:r_c}),
\beqa
\lefteqn{3000{\Omega _yh^2\over k_cv_c}
{50\hbox{ km s}^{-1}\over 1\hbox{ Mpc}}=} \label{eq:qc}\\
 & & \left[ 3\times 10^7\Omega _y^4 h^6\left( {hr_c\over v_c}
{50\hbox{ km s}^{-1}\over 1\hbox{ kpc}}\right) ^2\right] ^{(q-2)/4}. \nonumber
\eeqa
From the simulations in White \& Kroft (2000) it appears 
that if $k_c\gsim 10h\hbox{ Mpc}^{-1}$
the transfer function might be acceptable
for structure formation. At this bound, with 
$\Omega _y=0.25$ and $h=0.7$, and 
halo parameters $v_c=50$ km~s$^{-1}$ at
$r_c=1h^{-1}$~kpc, equation~(\ref{eq:qc}) says 
$q = 3.7$. Increasing the Jeans length to $hr_c = 3$~kpc
gives $q=3.4$. For $\Omega _y=1$ with the other parameters  
from the first example, $q=3.4$. If $q=3.6$ the physical Jeans
length varies quite slowly with the mass density,
$r_c\propto\rho _y^{-0.1}$. If $q=4$ the Jeans length is a
universal constant: $hr_c \lsim 0.5$~kpc at 
$k_c\gsim 10h\hbox{ Mpc}^{-1}$, $\Omega _y=0.25$, and $h=0.7$.

\section{Fluid Dark Matter}

The method of analysis for a pure quartic potential, and
numerical tests of the approximations, are presented in 
Peebles~(1999). The present case assumes the velocity of
sound and the dark matter 
streaming velocity are nonrelativistic. The former implies the
anharmonic part of $V(y)$ is subdominant, so the field is
oscillating at frequency $m$. If $q=4$ the equation of state is  
$p_y=3\lambda\rho _y^2/(8m^4)$, where
$K=\lambda$ (eq.~[\ref{eq:lambda}]), and 
\beq
m=\left( 9\lambda\over 8\pi Gr_c^2\right) ^{1/4} \sim
10^{16}\lambda ^{1/4}\left( 1\hbox{ kpc}\over r_c\right)^{1/2}
\hbox{ Hz}.
\label{eq:m}
\eeq
It seems safe to assume
this is much more rapid than the rate at which the mass
distribution and flux are changing, so we can find the 
mass density and velocity as functions of position and time by 
averaging functions of the field over a time interval $\tau$ such
that   
\beq
\dot\rho _y/\rho _y\ll\tau ^{-1}\ll m .
\label{eq:tau}
\eeq
Thus the mass density is well approximated as
\beq
\rho _y =\lb\dot y^2\rb = m^2\lb y^2\rb .
\label{eq:qrho}
\eeq
In the time and space intervals $\tau$ and $v\tau$ a good
approximation to a solution to the wave equation is   
\beq
y=\eta (\gamma (t-{\bf v}\cdot {\bf r})),
\label{eq:gamma}
\eeq
where $\ddot\eta =-dV(\eta )/d\eta$.
This is a Lorentz transformation
of $\eta (t)$, with 
Lorentz factor $\gamma$, so it describes proper mass density 
$\lb\dot \eta ^2\rb$ moving with velocity {\bf v}. Consistent
with this, the field gradient is
\beq
\nabla y = -\dot y{\bf v},
\label{eq:grad_y}
\eeq
for small $v$, and the mass flux density is
\beq
\lb {\bf f}_y\rb = -\lb\dot y\nabla y\rb = \rho _y{\bf v}.
\label{eq:fy}
\eeq

Consider first the pressure balance in a static dark matter halo. 
The line element is $ds^2 = (1+2\phi )dt^2 - (1 - 2\phi )dr^2$,
where the Newtonian potential $\phi$ is a function of position but
independent of time, and terms of order $\phi ^2$ dropped. 
The wave equation is
\beqa
\lefteqn{(1-4\phi )\ddot y -\nabla ^2y} \nonumber\\
& & + (1-2\phi )(m^2y +Ky|y|^{q-2}) = 0.
\label{eq:wave_eq}
\eeqa
The result of multiplying this equation by $y$ and averaging over
time, and using $\lb y\ddot y\rb =-\lb\dot y^2\rb$, from
integration by parts, is the virial relation 
\beqa
\lefteqn{(1-4\phi )\lb\dot y^2\rb +\lb y\nabla ^2y\rb } \nonumber\\
& & = (1-2\phi )(m^2\lb y^2\rb + K\lb |y|^q\rb ).
\label{eq:virial}
\eeqa
The result of multiplying equation~(\ref{eq:wave_eq}) by the gradient
of $y$ with respect to the Cartesian position component
$r_\alpha$, averaging over time, and using
equation~(\ref{eq:virial}), is 
\beqa
\lefteqn{\phi _{,\alpha }\lb m^2y^2 +K|y|^q - 2y\nabla ^2y\rb }\nonumber\\
& & + \partial _\beta (\lb y_{,\alpha }y_{,\beta }\rb - 
\partial _\alpha\partial _\beta\lb y^2\rb /4) \label{eq:static1}\\
& & +(1-2\phi )\partial _\alpha p_y = 0,\nonumber
\eeqa
where 
\beq
p_y = (q-2)K \lb |y|^q\rb /(2q).
\label{eq:p_y_p}
\eeq
The coefficient of  $\phi _{,\alpha}$ may be written 
as $m^2\lb y^2\rb =\rho _y$,
because the anharmonic term is subdominant in a nonrelativistic
fluid and equations~(\ref{eq:qrho}) and~(\ref{eq:grad_y}) 
say $\lb y\nabla ^2y\rb\sim\lb y\ddot yv^2\rb\sim
-\rho _yv^2$. The second term in the second
line is small because it depends on the astronomical length scale
of variation of the mean density. In last term the Newtonian
potential is an unimportant correction to the pressure gradient
force. Thus the balance condition is
\beq
\rho _y\phi _{,\alpha } + \p _\alpha p_y + 
\partial _\beta\lb\rho _y v_\alpha v_\beta\rb = 0.
\label{eq:static}
\eeq
The last term, which uses equation~(\ref{eq:grad_y}), is the
analog of the Reynolds stress in a turbulent fluid or the
velocity dispersion in a gas. Equation~(\ref{eq:p_y_p}) 
for $p_y$ is reduced to equation~(\ref{eq:p_y}) by using
$y\propto\sin mt$. 

When the Reynolds stress term may be neglected
equation~(\ref{eq:static}) yields the familiar relation between  
the halo central density, the core radius $r_c$, and the velocity
of sound in equation~(\ref{eq:rhoc}).

Consider next dynamics under the simplifying
assumption that the time-dependence of the Newtonian potential
$\phi$ may be ignored. The virial relation in
equation~(\ref{eq:virial}) applies here as an average over the
time and space intervals $\tau$ and $v\tau$
(eq.~[\ref{eq:tau}]). 
The result of using the wave equation to eliminate
$\ddot y$ from the time derivative of the mass flux density ${\bf
f}=-\dot f\nabla f$, averaging, and using the virial relation, is 
\beqa
\lefteqn{\partial \lb f_\alpha\rb /\partial t +
\phi _{,\alpha }\lb m^2y^2 + K|y|^q 
- 2y\nabla ^2y\rb } \nonumber\\
& & + (1+4\phi )\partial _\beta (\lb y_{,\alpha }y_{,\beta }\rb - 
\partial _\alpha\partial _\beta\lb y^2\rb /4) \nonumber\\
& & + (1+2\phi )\partial _\alpha p_y = 0.\label{eq:mom1}
\eeqa
This generalizes equation~(\ref{eq:static1}) to a dynamical
situation. As before, the nonrelativistic limit is 
\beq
\partial (\rho _yv_\alpha )/\partial t 
+\partial _\beta (\rho _y v_{\alpha }v_{\beta }) =
-\rho _y\phi _{,\alpha } -\partial _\alpha p_y.
\label{eq:mom}
\eeq
The mass density is
\beqa
\lefteqn{\rho _y=(1-4\phi )\dot y^2/2 + \nabla y^2/2 }\nonumber\\
& & + (1-2\phi )(m^2y^2/2 + K|y|^q/q),
\eeqa
with $\dot\rho _y=-\nabla\cdot {\bf f}=-\nabla\cdot\rho _y{\bf v}$.
This with equation~(\ref{eq:mom}) is the Euler equation 
--- the Navier-Stokes equation with zero viscosity --- for a
fluid with pressure $p_y$ in the gravitational field 
${\bf g}=-\nabla\phi$. 

Similar methods show the mean mass density in an expanding
universe with $\phi =0$ satisfies
\beq
\lb\dot\rho _y\rb = - 3{\dot a\over a}\left( \lb\rho _y\rb +
\lb p_y\rb + {\lb (\nabla y)^2\rb\over 3a^2}\right) .
\eeq
Here again the intrinsic pressure 
$p_y$ (eq.~[\ref{eq:p_y_p}]) is added to 
the contribution $\lb\rho _yv^2\rb /3$ by turbulence in a
fluid or velocity dispersion in a gas.

The ideal fluid picture cannot be complete: the velocity of sound is
an increasing function of $\rho _y$, so pressure waves tend to
grow into the analog of shock waves. The numerical examples in
Peebles (1999) for a pure  
quartic potential indicate that, unlike real shock waves, the
shock-like features tend to be 
reversible. This fluid picture with reversible ``shocks'' must 
eventually break down by the cascade of field energy to
wavelengths short enough that the better analog is a gas of
particles. To be investigated is the time for this to happen.  
This limitation of the fluid picture does not seem 
to be a problem for the halo model (eq.~[\ref{eq:static}]),
because it depends on the apparently reasonable assumption that
the $y$-field is a stationary time process. 

\section{Discussion}

Dark matter pictured as an interacting gas has been considered
by Carlson, Machacek, \&\ Hall (1992), Machacek (1994), and 
de Laix, Scherrer, \&\ Schaefer (1995). 
Spergel \&\ Steinhardt (1999) show interactions can have
interesting effects on the dark matter
distribution; the remarkable burst of papers on 
the concept includes Ostriker (1999); Hannestad (1999);
Miralda-Escud\'e (2000); Moore et
al. (2000); Hogan \&\ Dalcanton (2000); Yoshida et al. (2000);
Burkert (2000); and Firmani et al. (2000). 
This gas picture assumes small quantum occupation numbers. The
fluid picture discussed here and in PV assumes the opposite limit
of dark matter squeezed --- perhaps by inflation --- into a
state with large occupation numbers that behaves like a
classical field. The quadratic part of $V(y)$ (eq.~[\ref{eq:V(y)}]) 
describes particles of mass $m$ and present mean number density   
$\sim 100\lambda ^{-1/4}$~cm$^{-3}$ (eq.~[\ref{eq:m}]), but one
should think of these particles as moving coherently with large
de Broglie wavelength, as a classical field. The examples in \S 3
show this field can behave like an ideal fluid. 

The interacting gas picture may be approximated as a fluid, too,
but with two important differences. First, the Jeans length in a
patch of self-interacting gas depends on its history, while the
potential $V(y)$ fixes an intrinsic Jeans length 
(eq.~[\ref{eq:r_c}]). Second,
self-interacting gas acts as a viscous fluid, with ram pressure,
while the analysis here and in Peebles (1999) shows 
situations where the fluid dark matter has no viscosity  
(eq.~[\ref{eq:mom}]), and soliton-like behavior (Peebles 1999). 

Pressure in the fluid dark matter could suppress the small-scale
end of the spectrum of primeval fluctuations in the dark mass
distribution. Kam\-ion\-kowski \&\ Liddle (1999) suggest a
suppression could correct indications of excess numbers of dwarf
galaxies in the adiabatic cold dark matter model (Klypin et al.
1999 and references therein), and note the suppression could have 
originated during inflation. Warm dark matter 
(Hogan \&\ Dalcanton 2000) and fluid dark matter produce a
similar effect from a simpler inflation, at the price of more
complicated dark matter.  

The model for the potential (eq.~[\ref{eq:V(y)}])
is most elegant if the anharmonic part is quartic, $q=4$, so
$V(y)$ is an analytic function of $y$ that can directly relate to
a simple model for eternal inflation (eq.~[\ref{eq:lambda}]). In
this case natural choices for the other parameters require Jeans
length $hr_c\lsim 0.5$~kpc, which may be somewhat small for dark
matter halos. Halos can be supported at larger core
radii in the ordinary way, by Reynolds stress
(eq.~[\ref{eq:static}]), but this leaves open the issue of solid
body rotation curves. Since 
$V(y)$ is an effective potential for the classical field limit 
it might not be unreasonable to consider non-integer $q$. If 
$0<q-2\ll 2$ then $V(y)$ 
can produce dark matter halos with interesting solid
body rotation without seriously affecting the
primeval mass fluctuation spectrum. 
If $q\sim 3.5$ the model can have interesting effects on
both.

The fluid model requires the core radii $r_c$ of dark
halos satisfy equation~(\ref{eq:r_c}) as a lower bound; velocity
dispersion can only increase $r_c$ at   
given central density. Dalcanton \&\ Bernstein (1999) present
an elegant example of solid body rotation in a low surface
brightness galaxy that seems to be a good approximation to
a dark matter halo. The critical question, whether such behavior
is universal, has a long history and is under 
discussion (Salucci \&\ Persic 1999; van den Bosch et al. 1999;
Swaters, Madore, \&\ Trehwella 2000; Firmani et al. 2000; and
references therein).    

The survival of the dark matter in dwarf satellite galaxies is 
an important issue for the interacting gas picture 
(Moore et al. 2000) and the fluid picture. In the simple
cases in Peebles (1999) lumps in the fluid dark matter mimic
solitons. Fluid dark matter in the halos of dwarf
satellite galaxies might similarly pass through a larger dark
matter halo with little disturbance, but the idea has not been
considered in any detail.  

One may also judge models by elegance. Thus one might
seek a simple reason why the present ratio 
$\rho _m/\rho _r$ of mass densities in dark matter and radiation
is not far from unity, compared to the great difference at
the end of inflation if the dark matter were quite cold and
noninteracting. This also has a long history,  
as in Lee \&\ Weinberg (1977); a recent proposal by Zlatev \&\
Steinhardt (1999) is another model of interacting dark matter,
with potential $V(y)$ such that the present value of 
$\rho _m/\rho _r$ is insensitive to a broad range of initial
conditions. The potential in equation~(\ref{eq:V(y)}) does not
exhibit this behavior but does allow another
consideration. It is easy to imagine the field that acted as
the inflaton during inflation had energy density only a few
orders of magnitude larger than some other field
(Kofman \&\ Linde 1987), and that the latter field 
happens to interact only with itself and gravity, 
so it ends up as dark matter (PV). The 
quartic form of $V(y)$ at large field values would keep
$\rho _m/\rho _r$ close to unity. 
The potential must change to  
near quadratic at the rms value of $y$ reached some
twenty orders of magnitude of expansion after inflation ends, in 
time to make the dark mass density comparable to the baryons. 
This still is a considerable ``cosmic coincidence.'' 

The bound on $\lambda$ 
(eq.~[\ref{eq:lambda}]) at high redshift depends on the highly
uncertain efficiency of production of 
ordinary matter and radiation as inflation ends. Within this
uncertainty $\lambda$ may be adjusted so structure formation is 
seeded by the curvature fluctuations from the inflaton, as in the
adiabatic cold dark matter model. The replacement of cold with
fluid dark matter in this model would affect structure formation
only on relatively small scales, where it might be beneficial.
Alternatively, one may postulate a larger value of $\lambda$ that
would produce significant isocurvature fluctuations, perhaps in a
more elaborate version of the model in Hu \&\ Peebles (1999).
This could be beneficial if the model assembled galaxies
earlier than predicted in the cold dark matter model, and earlier
formation better fit the observations. 

\acknowledgements

I have benefitted from discussions with
Julianne Dalcanton, Paul Steinhardt and Alex Vilenkin. 
This work was supported in part by the NSF.

\end{document}